\RequirePackage{lineno}
\documentclass[showpacs,aps,amsmath,amssymb,prc,showpacs,superscriptaddress,floatfix,preprint,preprintnumbers]{revtex4-1}
\usepackage{rotating}
\usepackage{epsfig}
\usepackage[colorlinks=true, allcolors=blue]{hyperref}
\usepackage{lineno}

\usepackage{amsmath}
\usepackage{graphicx}
\usepackage[caption=false]{subfig}

\usepackage[letterpaper,top=2cm,bottom=2cm,left=3cm,right=3cm,marginparwidth=1.75cm]{geometry}

\usepackage[dvipsnames]{xcolor}

\newcommand\snn{\sqrt{s_{_{NN}}}}
\newcommand\gevc{GeV/$c$}
\newcommand\rp{{\rm p}}
\newcommand\rn{{\rm n}}
\newcommand\rd{{\rm d}}
\newcommand\rt{{\rm t}}
\newcommand\Ratio{N_\rt N_\rp/N_\rd^2}

\newcommand\pt{\ensuremath{p_{\rm{T}}}}
\newcommand\mt{\ensuremath{m_{\rm{T}}}}

\newcommand\he{\ensuremath{^3{\rm He}}}

\newcommand\Nch{N_{\rm ch}}

\newcommand\MEAN[1]{\langle#1\rangle}

\begin{document}

\title{Acceptance effect on the $\Ratio$ ratio of light nuclei coalescence yields as a probe of nucleon density  fluctuations}

\author{Michael X.~Zhang}
\affiliation{Livermore High School, Livermore, CA 94550, USA}
\author{An Gu}
\email{Corresponding author: gu180@purdue.edu}
\affiliation{Purdue University, West Lafayette, IN 47907, USA}
\affiliation{Huzhou University, Huzhou 313000, China}

\begin{abstract}
We employ a coalescence model to form deuterons (\rd), tritons (\rt) and helium-3 (\he) nuclei from a uniformly distributed volume of protons (\rp) and neutrons (\rn). 
We study the ratio $\Ratio$ of light nuclei yields as a function of the neutron density fluctuations. 
We investigate the effect of finite transverse momentum ($\pt$) acceptance on the ratio, in particular, the ``extrapolation factor" ($f$) for the ratio as functions of the $\pt$ spectral shape and the magnitude of neutron density fluctuations. 
It is found that $f$ is monotonic in $\pt$ spectra ``temperature" parameter and neutron density fluctuation magnitude; variations in the latter are relatively small. 
We also examine $f$ in realistic simulations using kinematic distributions of protons measured in heavy ion collision data. It is found that $f$ is smooth and monotonic as a function of beam energy. We conclude that extrapolation from limited $\pt$ ranges would not create, enhance, or reduce a local peak of the  $\Ratio$ ratio in beam energy.
Our study provides a necessary benchmark for light nuclei ratios as a probe of nucleon density fluctuations, an important observable in the search for the critical point of nuclear matter. 
\end{abstract}

\maketitle

\section{Introduction}
Matter is comprised of quarks and gluons, the most fundamental constituents in nature together with leptons and gauge bosons. The interactions among quarks and gluons are governed by quantum chromodynamics (QCD). At low temperature and matter density quarks and gluons are confined in hadrons, while at high temperature or matter density they are deconfined in an extended volume called the quark-gluon plasma (QGP). The phase transition at low temperature and high matter density is of first order, and at high temperature and low matter density it is a smooth crossover as predicted by lattice QCD~\cite{Aoki:2006we}. It has been conjectured that a critical point (CP) exists on the nuclear matter phase diagram of temperature versus matter density between the first-order phase transition and the smooth crossover~\cite{Halasz:1998qr,Stephanov:1998dy,Stephanov:2004wx,Fukushima:2010bq}. The correlation length increases dramatically near the CP causing large fluctuations of conserved quantities, such as the net-baryon number~\cite{Stephanov:2011pb}. 
Searching for the CP is a subject of active research in heavy-ion collisions~\cite{STAR:2010vob,STAR:2013gus,Luo:2015doi,Bzdak:2019pkr,ALICE:2019nbs,HADES:2020wpc,STAR:2020tga,STAR:2021iop,STAR:2021fge}.

Light nuclei production is well modeled by nucleon coalescence~\cite{Butler:1963pp,Sato:1981ez,Csernai:1986qf,Dover:1991zn,Scheibl:1998tk,Chen:2003qj,Oh:2009gx}. 
It has been predicted by the coalescence model that large net-baryon number fluctuations affect the production rate of light nuclei~\cite{Sun:2017xrx,Sun:2018jhg}. For example, the production of tritons (\rt) would be enhanced relative to that of deuterons (\rd) when there are extra fluctuations in the neutron density because a triton contains two neutrons (\rn) whereas a deuteron contains only one. As a result, the compound ratio of $\Ratio$ involving the multiplicities of protons ($N_\rp$), deuterons ($N_\rd$) and tritons ($N_\rt$) would be enhanced. Similarly for the production of helium-3 ($\he$) and the ratio of $N_{\he} N_\rn/N_\rd^2$ with respect to extra fluctuations in the proton density.

Following Ref.~\cite{Sun:2017xrx,Sun:2018jhg,Sun:2017ooe}, the deuteron average multiplicity density in the coalescence model is given by
\begin{equation}
    \bar{n}_\rd/(\frac{3}{\sqrt{2}}A)=\MEAN{(\bar{n}_\rp+\delta n_\rp)(\bar{n}_\rn+\delta n_\rn)}
    =\bar{n}_\rp\bar{n}_\rn+\MEAN{\delta n_\rp\delta n_\rn}
    =\bar{n}_\rp\bar{n}_\rn(1+(\alpha\Delta n_\rn)_\rd)\,,
    \label{eq:d}
\end{equation}
where the protons and neutrons are assumed to be in thermal equilibrium with an  effective temperature $T_{\rm eff}$ and $A=(\frac{2\pi}{m_N T_{\rm eff}})^{3/2}$ is a shorthand notation ($m_N$ is the nucleon mass). Here and henceforth, $\bar{n}$ stands for average density. The neutron density fluctuations are denoted by
\begin{equation}
    \Delta n_\rn=\MEAN{(\delta n_\rn)^2}/\bar{n}_\rn^2\,,
    \label{eq:Dn}
\end{equation}
and $\alpha=\frac{\MEAN{\delta n_\rp\delta n_\rn}}{\bar{n}_\rp\bar{n}_\rn}/\Delta n_\rn=\frac{\bar{n}_\rn}{\bar{n}_\rp}\frac{\MEAN{\delta n_\rp\delta n_\rn}}{\MEAN{(\delta n_\rn)^2}}$ denotes the correlations between proton and neutron number fluctuations. If proton and neutron numbers fluctuate independently, then $\alpha=0$; if they fluctuate together, then $\alpha=1$. 
It is noteworthy at this point that we have so far neglected details of deuteron formation which is determined by the deuteron wavefunction and often implemented by the Wigner function formalism (see below). Fluctuations that matter for deuteron formation are those within typical volumes of the size of the deuteron. We note this in Eq.~(\ref{eq:d}) by the subscript `\rd' in $(\alpha\Delta n_\rn)_\rd$ to indicate that it is the average $\alpha\Delta n_\rn$ within the typical deuteron size that is relevant.

Similarly, the triton average multiplicity density is given by
\begin{equation}
    \bar{n}_\rt/(\frac{3\sqrt{3}}{4}A^2)=\MEAN{(\bar{n}_\rp+\delta n_\rp)(\bar{n}_\rn+\delta n_\rn)^2}
    =\bar{n}_\rp\bar{n}_\rn^2(1+(\Delta n_\rn)_\rt+2(\alpha\Delta n_\rn)_\rt+(\beta\Delta n'_\rn)_\rt)\,,
    \label{eq:t}
\end{equation}
where $\Delta n'_\rn=\MEAN{(\delta n_\rn)^3}/\bar{n}_\rn^3$ and $\beta=\frac{\MEAN{\delta n_\rp(\delta n_\rn)^2}}{\bar{n}_\rp\bar{n}_\rn^2}/\Delta n'_\rn=\frac{\bar{n}_\rn}{\bar{n}_\rp}\frac{\MEAN{\delta n_\rp(\delta n_\rn)^2}}{\MEAN{(\delta n_\rn)^3}}$. 
Similar to the $\alpha$ in Eq.~(\ref{eq:d}), the $\beta$ parameters denotes three-body correlations. When proton and neutron numbers fluctuate independently, then $\beta=0$; when they fluctuate together, then $\beta=1$.
The subscript `\rt' in Eq.~(\ref{eq:t}) likewise indicates that only those fluctuations averaged within the typical volume of the triton size matter. Note that in Eq.~(\ref{eq:t}) we simply wrote $n^2_\rn$ for neutron pair density, however, for identical particles pair multiplicity is $N(N-1)$, so the pair fluctuations of Eq.~(\ref{eq:Dn}) should be understood as those beyond Poisson fluctuations.

The compound ratio is thus given by
\begin{equation}
    \frac{N_\rp N_\rt}{N_\rd^2}=\frac{\bar{n}_\rp\bar{n}_\rt}{\bar{n}_\rd^2}=\frac{1+(\Delta n_\rn)_\rt+2(\alpha\Delta n_\rn)_\rt+(\beta\Delta n'_\rn)_\rt}{2\sqrt{3}(1+(\alpha\Delta n_\rn)_\rd)^2}
    \,.
    \label{eq:r}
\end{equation}
%
%
If one neglects $\alpha$ and $\beta$, then Eq.~(\ref{eq:r}) reduces to 
\begin{equation}
    \frac{N_\rp N_\rt}{N_\rd^2}\approx\frac{1}{2\sqrt{3}}(1+(\Delta n_\rn)_\rt)\,,
    \label{eq:simple}
\end{equation}
as in Ref.~\cite{Sun:2017xrx,Sun:2018jhg}. Note that the factor $1/2\sqrt{3}$ comes from the thermal equilibrium assumption of nucleon abundances.
The $\Ratio$ ratio may therefore be a good measure of the neutron density fluctuations, a large value of which could signal the CP.

\begin{figure}[hbt]
    \includegraphics[width=0.48\linewidth]{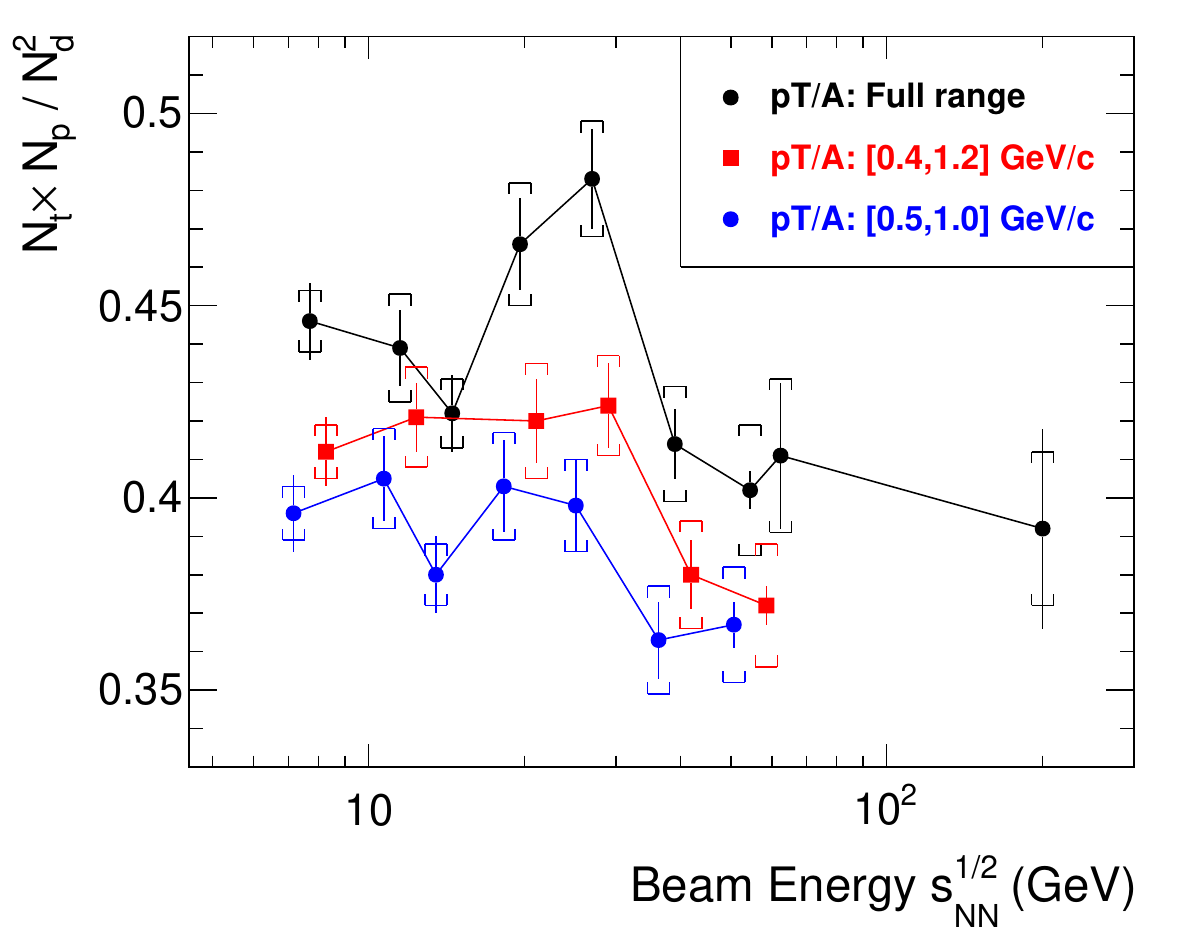}\hfill
    \includegraphics[width=0.48\linewidth]{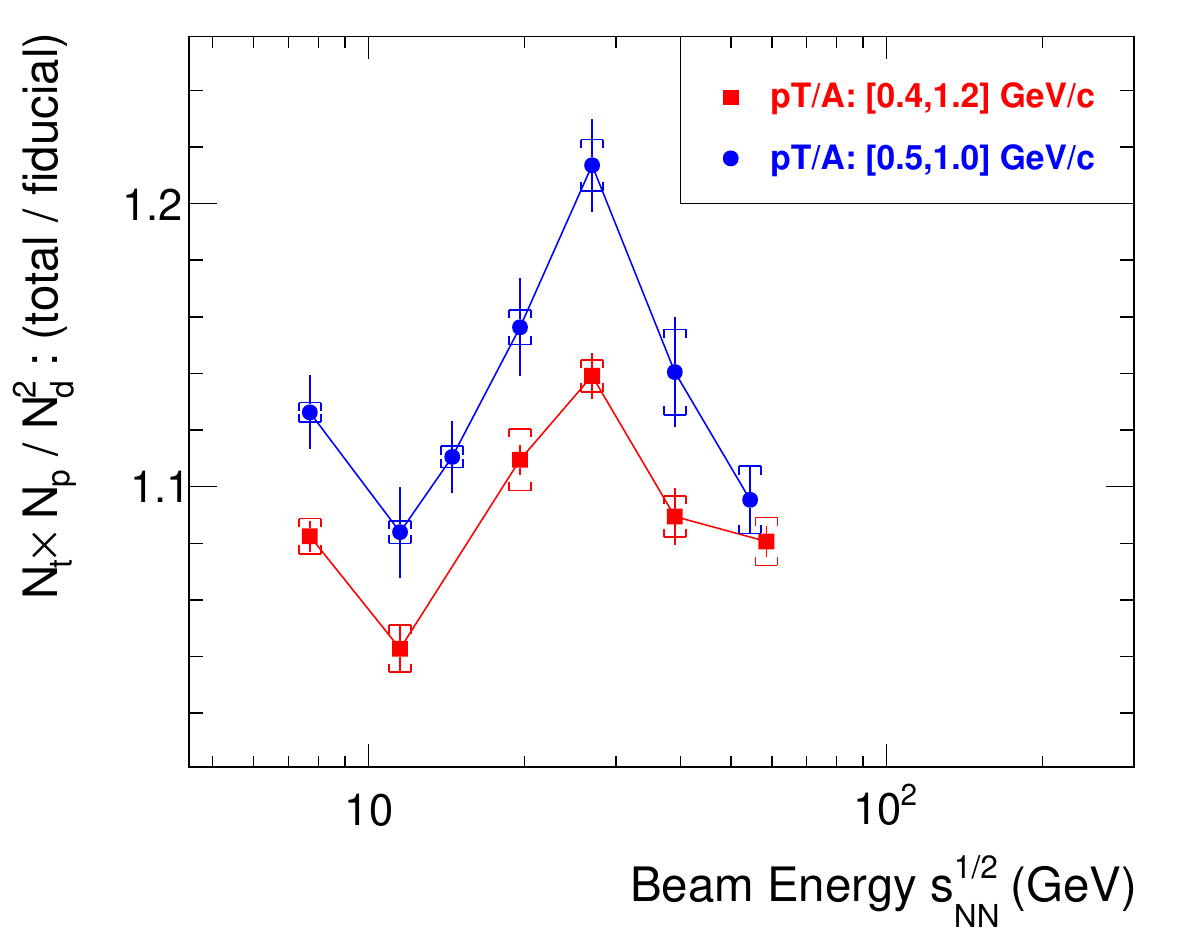}
    \caption{\label{fig:STAR}(Left) The $\Ratio$ ratio in 0-10\% central Au+Au collisions as functions of beam energy $\snn$ measured by STAR~\cite{STAR:2022hbp}. The ratio of the extrapolated total yields and those from two measured $\pt/A$ ranges are shown. The ratios from the measured $\pt/A$ ranges are shifted in the horizontal axis for clarity.
    (Right) The extrapolation factor $f$, i.e., the $\Ratio$ ratio from the extrapolated total yields divided by that from the fiducial yields in a given measured $\pt/A$ range, is shown for the two measured $\pt/A$ ranges as functions of $\snn$. The statistical and systematic uncertainties on $f$ are the quadratic {\em difference} of the corresponding uncertainties of the $\Ratio$ ratio between the given measured $\pt/A$ range and the total range. The rightmost square is shifted in the horizontal axis for clarity.}
\end{figure}
A unique signature of the CP would be a non-monotonic behavior of the ratio $\Ratio$ in beam energy, where a peak of the ratio in a  localized region of beam energy could signal large neutron fluctuations and the CP~\cite{Sun:2017xrx,Sun:2018jhg}. The STAR experiment at RHIC has indeed  observed recently a non-monotonic behavior of the $\Ratio$ ratio in top 10\% central Au+Au collisions as a function of the nucleon-nucleon center-of-mass energy ($\snn$) in the Beam Energy Scan (BES) data~\cite{STAR:2022hbp}. 
The non-monotonic bump is observed in the ratio localized in the energy region $\snn=20$--30~GeV. It is noted that the local bump is prominent in the $\Ratio$ ratio of the extrapolated yields to all transverse momenta ($\pt$), but not as prominent in the measured fiducial range of $0.5<\pt/A<1.0$~\gevc\ and $0.4<\pt/A<1.2$~\gevc\ (where $A$ is the mass number corresponding to each light nucleus in the ratio)~\cite{STAR:2022hbp}.
This is shown in the left panel of Fig.~\ref{fig:STAR} where the STAR measured $\Ratio$ ratios from the total extrapolated yields and from the two measured $\pt/A$ ranges are reproduced. 
The ``extrapolation factor" ($f$), i.e., the  $\Ratio$ ratio of the total light nuclei yields extrapolated to the entire $\pt/A$ range  [0--$\infty$) divided by the ratio of those measured within a fiducial $\pt/A$ range, is shown in the right panel of Fig.~\ref{fig:STAR} for the two measured $\pt/A$ ranges. The statistical uncertainties between the fiducial yield of a given $\pt$ spectrum and the extrapolated total yield are correlated, so are the systematic uncertainties. Thus, the statistical and systematic uncertainties are taken to be the quadratic {\em difference} of the corresponding uncertainties on the $\Ratio$ ratio between the fiducial $\pt/A$ range and the total range. As expected from the STAR results~\cite{STAR:2022hbp}, the $f$ values are peaked at the $\snn=20$--30~GeV range and are non-monotonic; the non-monotonic effect is on the order of 10\%.

In this paper, we use a toy model to generate nucleons and form \rd, \rt, and $\he$ using a coalescence model. We study the ratios of the light nuclei yields as functions of the magnitude of neutron multiplicity fluctuations, and examine the role the $\pt$ acceptance plays in those ratios. 
We first confine ourselves within a simple toy model to gain insights. We then try to carry out more realistic simulations mimicking the STAR BES data to examine what the STAR measurements may entail.

\section{Coalescence model}
The probability to form a composite particle from particle 1 at position $\vec{r}_1$ and with momentum $\vec{p}_1$ and particle 2 at position $\vec{r}_2$ and with momentum $\vec{p}_2$ is estimated by the Wigner function, 
\begin{equation}
W(\vec{r}_1, \vec{p}_1, \vec{r}_2, \vec{p}_2)=g\cdot 8\exp\left(-\frac{r_{12}^2}{\sigma_r^2}-\frac{p_{12}^2}{\sigma_p^2}\right)\,,
\label{eq:coal}
\end{equation}
where
\begin{eqnarray}
r_{12}&=&|\vec{r_{1}}-\vec{r_{2}}|\,,\nonumber\\
p_{12}&=&\mu\left|\frac{\vec{p_1}}{m_1}-\frac{\vec{p_2}}{m_2}\right|\,,\nonumber
\end{eqnarray}
%
and $\mu=\frac{m_1 m_2}{m_1+m_2}$ is the reduced mass of the two-body system. 
The parameter $\sigma_r$ is the characteristic coalescence size in configuration space and $\sigma_\rp=1/\sigma_r$ (where $\hbar=c=1$) is that in momentum space. 

From the Wigner function of Eq.~(\ref{eq:coal}), the root mean square (RMS) radius of the coalesced composite particle can be calculated as $R=\frac{\sqrt{3m_1m_2/2}}{m_1+m_2}\sigma_r$. The coalescence parameter $\sigma_r$ can therefore be determined by the particle size as
\begin{equation}
    \sigma_r=\frac{m_1+m_2}{\sqrt{3m_1m_2/2}}R\,.
\end{equation}

The deuteron is coalesced by a proton and a neutron. The RMS size of the deuteron is $R_\rd=1.96$~fm~\cite{Ropke:2008qk}, so for deuteron  $\sigma_r=\sqrt{\frac{8}{3}}R_\rd=3.20$~fm.
The triton (helium-3) is formed by coalescence between a deuteron and a neutron (proton), following the same prescription of Eq.~(\ref{eq:coal}). 
The triton RMS size is $R_\rt=1.59$~fm~\cite{Ropke:2008qk}, so for triton  $\sigma_r=\sqrt{3}R_\rt=2.75$~fm.
For $\he$ we also take $\sigma_r=2.75$~fm to be symmetric in our study.

The $g$-factor represents the probability to have the proper total spin of the composite particle, 
\begin{equation}
    g=(2S+1)/\Pi_{i=1}^2 \left(2s_i+1 \right)\,,
\end{equation}
where $S$ is the spin of the composite particle (1 for \rd, 1/2 for both \rt\ and \he), $s_i$ is the spin of each coalescing particle. 
For \rd\ formed from \rp\ and \rn, the $g$-factor is $3/4$. For \rt\ ($\he$) formed from \rd\ and \rn\ (\rp), it is $1/3$.




\section{Toy-Model Simulation Details\label{sec}}
We generate a system of nucleons with a toy model.
The nucleons are assumed to be uniformly distributed in a cylinder of 10~fm length and 10~fm radius. 
The transverse momentum spectra of the nucleons are assumed to be 
\begin{equation}
    dN/d\pt\propto\pt\exp(-\pt/T)\,,
    \label{eq:pt}
\end{equation}
where the ``temperature" parameter is set to $T=150$~MeV. Note that Eq.~(\ref{eq:pt}) is not a thermal distribution for massive particles; we use the term ``temperature" for convenience. The azimuthal angle of the momentum vector is uniformly distributed between 0 and $2\pi$. The pseudorapidity $\eta$ is assumed to be uniform between $-1$ and $1$. Whether to use rapidity or pseudorapidity is insignificant.

In this study, we do not use the thermal model to predict the average multiplicity, but rather set the average multiplicities of protons and neutrons both to $\bar{N}_\rp=\bar{N}_\rn=20$. 
This is simply for convenience since our goal is only to study the effect of neutron multiplicity fluctuations. The baseline of the $\Ratio$ ratio is therefore not the $1/2\sqrt{3}$, but rather is determined by the ratio of the degeneracy $g$ factors as $\frac{1}{3}\cdot\frac{3}{4} / (\frac{3}{4})^2 = \frac{4}{9}$~\cite{Wu:2022cbh}.

We assign Poisson fluctuations to the number of protons, and only vary the fluctuation magnitude for the number of neutrons (and focus on the compound ratio of $\Ratio$). The latter is achieved by negative binomial distributions, $P(N_\rn=k)=C^{k+r-1}_{k}(1-p)^kp^r$, where $r$ and $p$ are free parameters. 
The mean and variance are $\bar{N}_\rn=\frac{r(1-p)}{p}$ and $\sigma^2_{N_\rn}=\frac{r(1-p)}{p^2}$, respectively.
The fluctuation magnitude is given by $\sigma^2_{N_\rn}=\bar{N}_\rn/p\equiv\theta\bar{N}_\rn$ (where $\theta\equiv1/p\ge1$), which is always larger than Poisson fluctuations unless the probability $p=1$ when the negative binomial distribution reduces to Poisson. We use $p$ to control the magnitude of fluctuations in $N_\rn$ and choose the proper $r$ value to have the desired average neutron multiplicity  $\bar{N}_\rn$.

It is worthy to note that we utilize fluctuations in the total number of neutrons $N_\rn$ event-by-event to mimic fluctuations in the local neutron number density. The purpose of our study is to investigate the behavior of  $\Ratio$  as a function of the neutron density fluctuation magnitude, but not to suggest that density fluctuations must be caused by fluctuations in the total multiplicity. In our simulation, given a total $N_\rn$ in an event, neutrons are randomly distributed in the cylinder volume. Fluctuations in $N_\rn$ thus determines the magnitude of neutron density fluctuations which is an average over the entire volume.
In other words, the  $\Delta n_\rn$ in Eq.~(\ref{eq:Dn}) can be quantified by the fluctuations in $N_\rn$ beyond Poisson,
\begin{equation}
    \Delta n_\rn = (\sigma_{N_\rn}^2/\bar{N}_\rn-1)/\bar{N}_\rn 
    \equiv (\theta-1)/\bar{N}_\rn\,.
    \label{eq:dn}
\end{equation}
Note that $\theta$ quantifies the fluctuations in $N_\rn$ in the unit of Poisson fluctuations and $\theta-1$ quantifies that beyond Poisson fluctuations.

A total of $1.6 \times 10^8$ events are simulated.
Deuterons, tritons, and helium-3 are formed by coalescence. For each event, deuteron formation is first considered via double loops over protons and neutrons. 
For each deuteron, the formation of \rt\ and $\he$ are implemented by looping over the remaining nucleons. 
The nucleons have been reordered randomly so the probabilities to form \rt\ and \he\ are unbiased. 
A random number uniformly distributed between 0 and 1 is used to determine whether two particles coalesce into a light nucleus or not. If the random number is smaller than the Wigner function value in consideration, then the corresponding light nucleus is formed. Once a light nucleus is formed, the coalescing particles are removed from further consideration for coalescence. 

\section{Toy-Model Simulation Results and Discussions}

\begin{figure}[hbt]
    \includegraphics[width=0.48\linewidth]{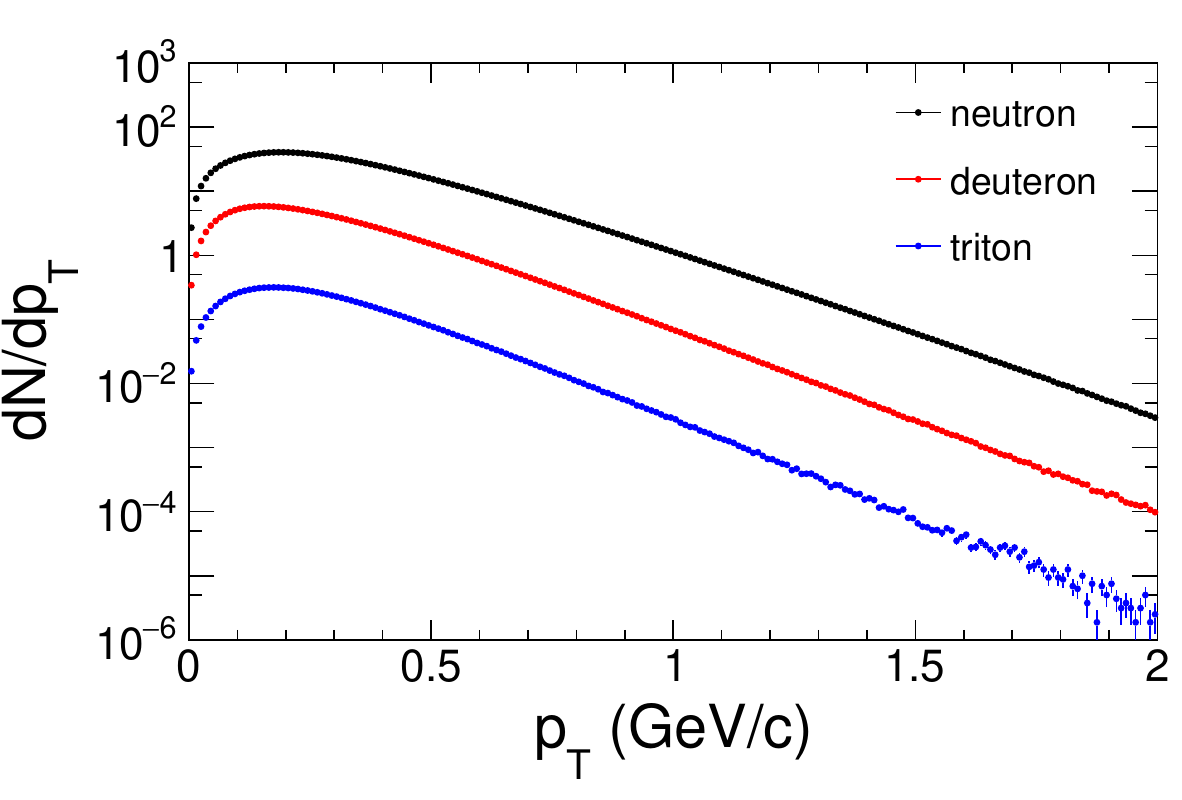}\hfill
    \includegraphics[width=0.48\linewidth]{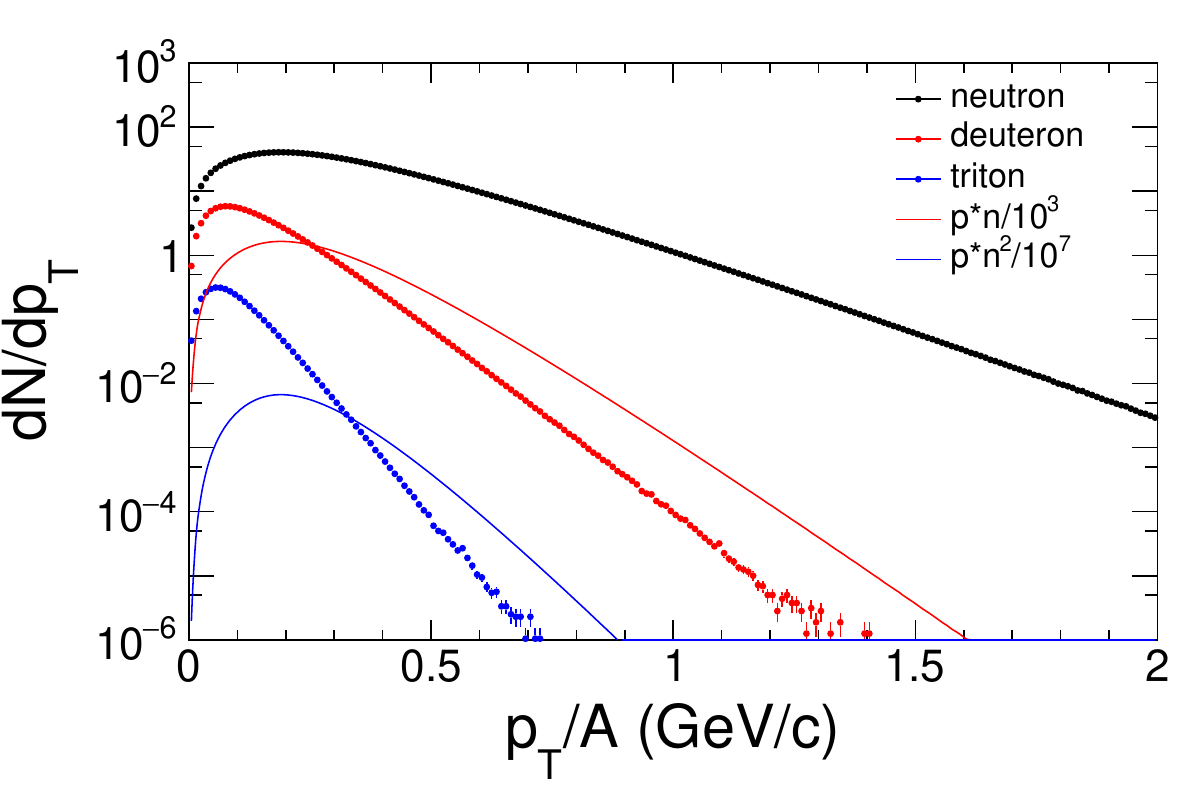}
    \caption{(Left) Transverse momentum $\pt$ spectra of deuteron and triton calculated by the coalescence model from a  system of average $\bar{N}_\rp=20$ protons and $\bar{N}_\rn=20$ neutrons at ``temperature" $T=150$~MeV (Eq.~(\ref{eq:pt})) randomly distributed within a cylinder of 10 fm radius and 10 fm length. No extra fluctuations are included beyond Poisson, i.e., $\theta=1$. (Right) The same spectra plotted as functions of $\pt/A$ (A is the corresponding mass number). Superimposed in curves are the products of $(dN_\rp/d\pt)\times(dN_\rn/d\pt)$ and $(dN_\rp/d\pt)\times(dN_\rn/d\pt)^2$ at the same $\pt/A$ value, arbitrarily scaled to compare to the shapes of the deuteron and triton spectra, respectively.} 
    \label{fig:spec}
\end{figure}

Figure~\ref{fig:spec} shows the $\pt$ spectra of the generated neutrons and the coalesced \rd, \rt, and $\he$ in the left panel, and in the right panel those spectra in the scaled $\pt/A$. No extra fluctuations are included beyond Poisson in this figure (i.e., $\theta=1$) so the proton spectrum is identical to that of neutrons. In the right panel, the product of the proton and neutron spectra and that of the proton and squared neutron spectra are shown in the smooth histograms. Those products would be the corresponding \rd\ and \rt\ spectra if the coalescence parameters $\sigma_p=0$ and $\sigma_r=\infty$ in Eq.~(\ref{eq:coal}). The coalesced \rd\ and \rt\ spectra are steeper than the products because of the finite $\sigma_\rp$ and $\sigma_r$ implemented in our coalescence model.

\begin{figure}[hbt]
\includegraphics[width=0.48\linewidth]{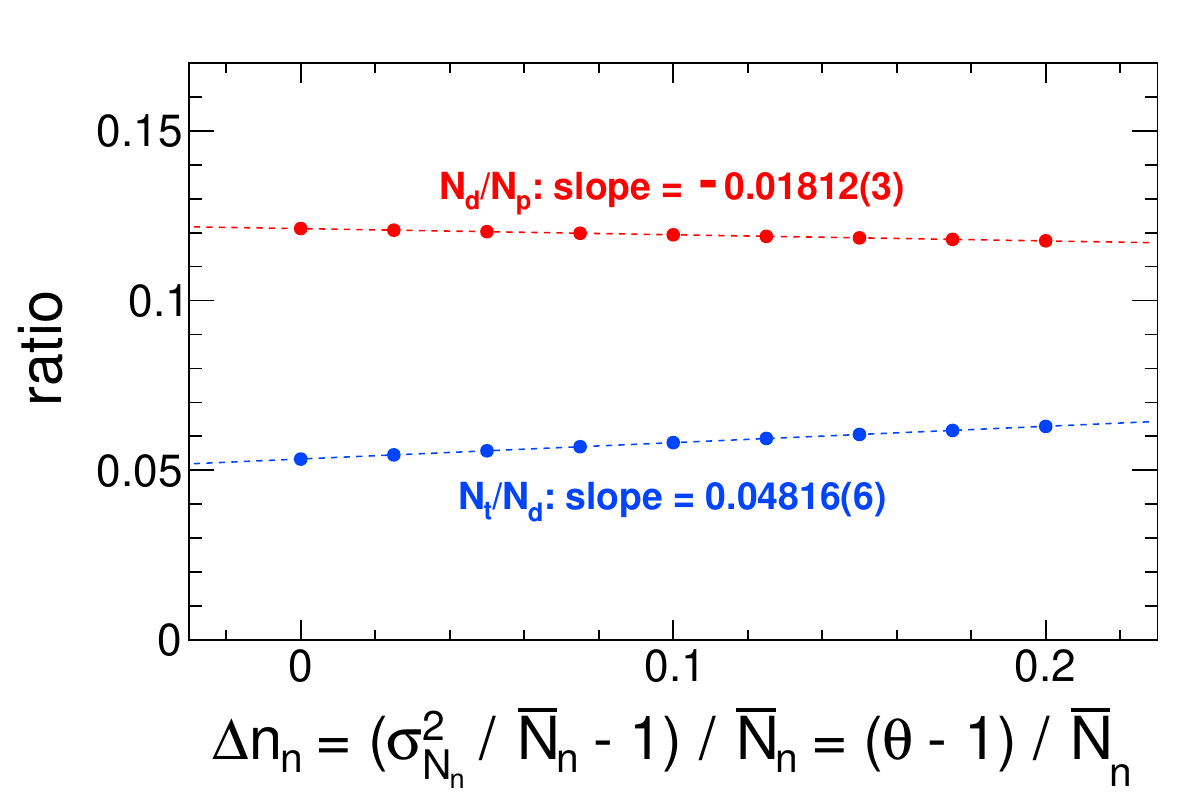}\hfill
\includegraphics[width=0.48\linewidth]{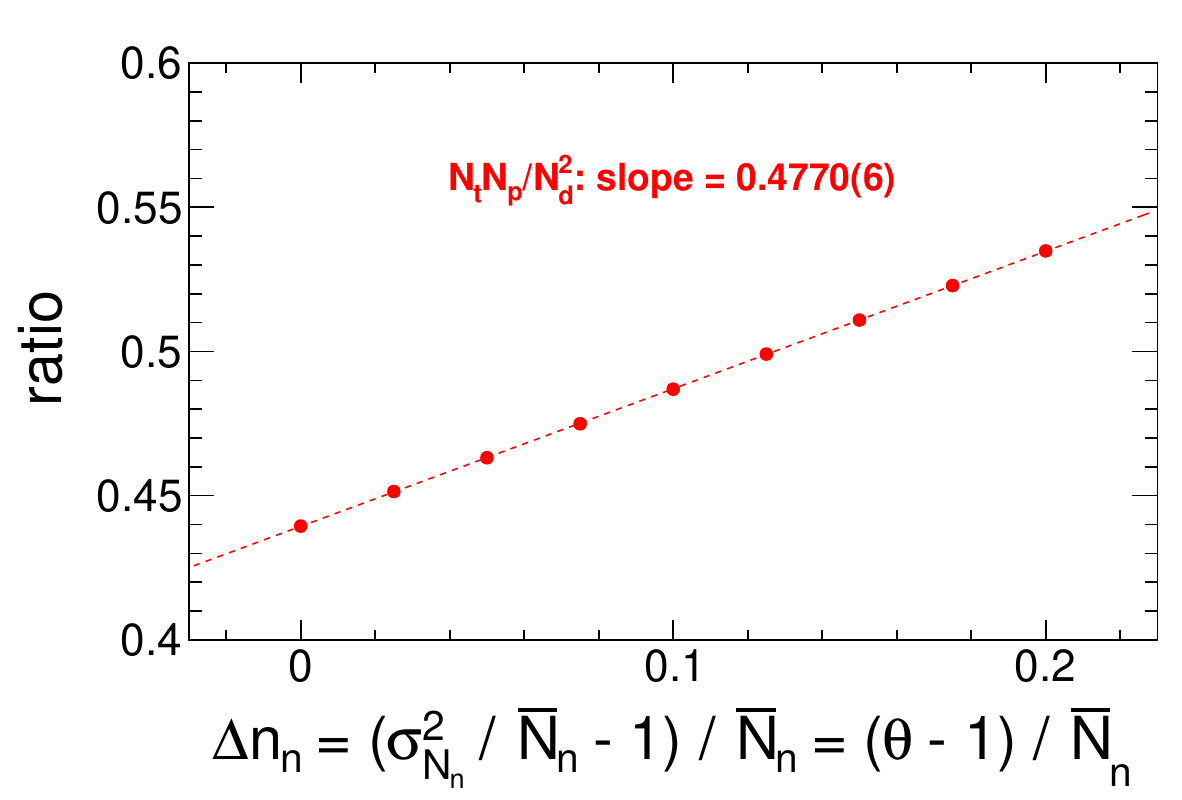}
    \caption{Yield ratios of $N_\rd/N_\rp$ and $N_\rt/N_\rd$ (left) and $\Ratio$  (right) as functions of $\Delta n_\rn$ of Eq.~(\ref{eq:dn}), the magnitude of neutron multiplicity fluctuations beyond Poisson. The light nuclei are formed by coalescence  from a  system of average $\bar{N}_\rp=20$ protons and $\bar{N}_\rn=20$ neutrons at $T=150$~MeV (Eq.~(\ref{eq:pt})) randomly distributed within a cylinder of 10 fm radius and 10 fm length.}
    \label{fig:ratio}
\end{figure}

Figure~\ref{fig:ratio} shows in the left panel the yield ratios of $N_\rd/N_\rp$ and $N_\rt/N_\rd$ as functions of the neutron fluctuation magnitude $\Delta n_\rn$ of Eq.~(\ref{eq:dn}). The $N_\rt/N_\rd$ ratio increases with $\Delta n_\rn$, consistent with the expected stronger effect of neutron density fluctuations on $\rt$ than on $\rd$ production. 
The $N_\rd/N_\rp$ ratio decreases slightly but statistically significant. This is counter-intuitive as one would expect no dependence because $\bar{N}_\rn$ is the same for all values of $\Delta n_\rn$; the events with more neutrons would be balanced out by those with fewer neutrons in terms of deuteron production.
However, $\bar{N}_\rp=20$ is fixed and fluctuations in $N_\rp$ are treated uncorrelated to those in $N_\rn$ in our study. The production of deuterons will be ``saturated" in events with large $N_\rn$--those neutrons would not get ``equal" share of protons to form deuterons. As a result, the $N_\rd/N_\rp$ is smaller for larger $\Delta n_\rn$.

The right panel of Fig.~\ref{fig:ratio} shows the $\Ratio$ ratio as a function of $\Delta n_\rn$. 
The $\Ratio$ ratio clearly increases with $\Delta n_\rn$. The slope is about 0.477, 
significantly non-zero, and the intercept is about 0.439. 
These values are roughly equal to the expected degeneracy factor 4/9. The slight deviations may be due to the different size parameters for deuteron and triton.

\begin{figure}[hbt]
    \includegraphics[width=0.48\linewidth]{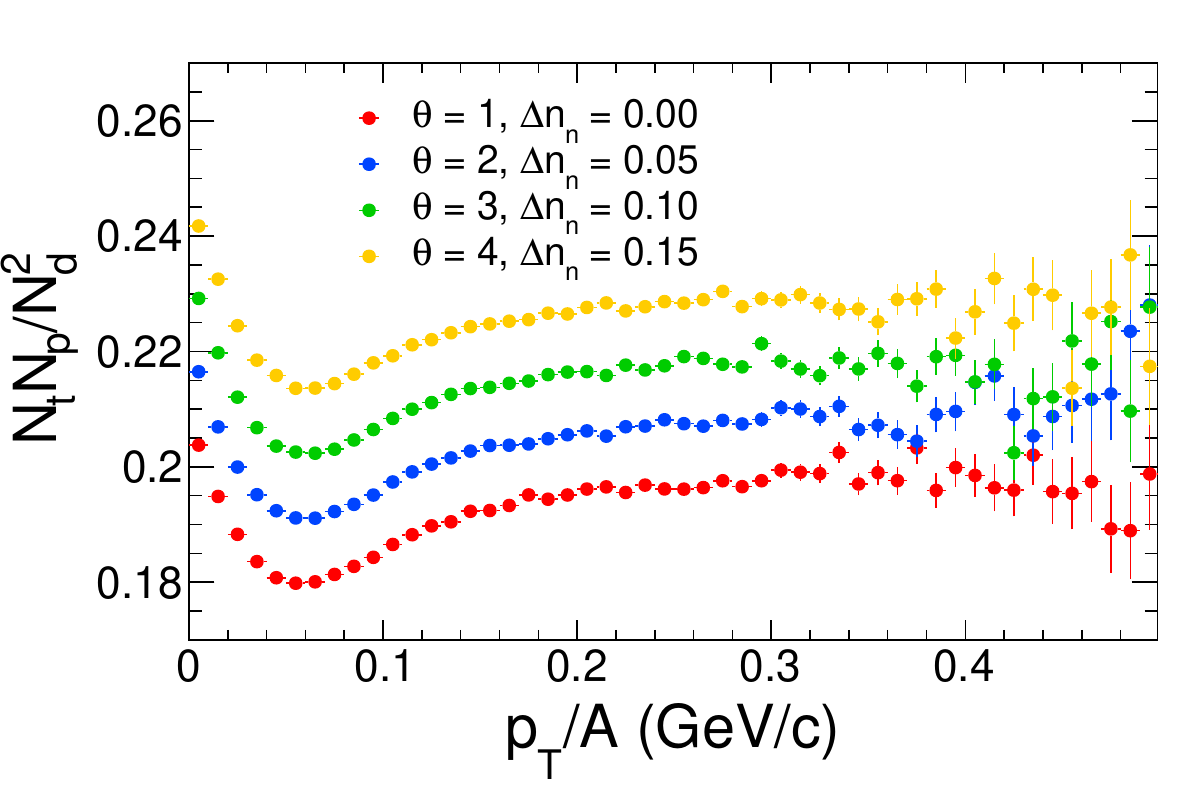}\hfill
    \includegraphics[width=0.48\linewidth]{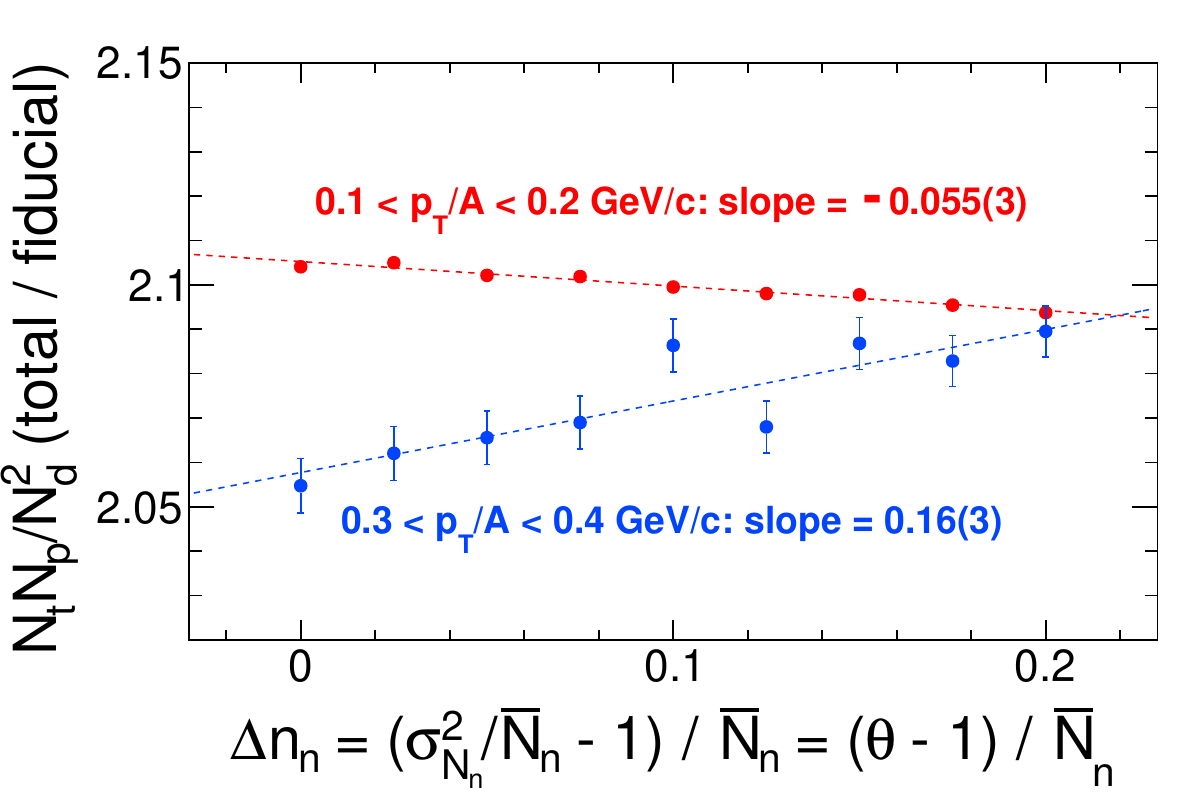}
    \caption{\label{fig:acc}(Left) The $\Ratio$ ratio as functions of $\pt/A$. (Right) The $\Ratio$ extrapolation factor $f$ from $\pt/A=0.1$--0.2~\gevc\ and 0.3--0.4~\gevc\ as a function of $\Delta n_\rn$. The light nuclei are formed by coalescence from a  system of average $\bar{N}_\rp=20$ protons and $\bar{N}_\rn=20$ neutrons at $T=150$~MeV (Eq.~(\ref{eq:pt})) randomly distributed within a cylinder of 10 fm radius and 10 fm length.} 
\end{figure}

As aforementioned, the measured non-monotonic feature of the $\Ratio$ ratio as a function of beam energy $\snn$ by STAR appears to depend on the considered $\pt/A$ range~\cite{STAR:2022hbp}; the non-monotonic feature is  stronger for the extrapolated yield ratio than for those in the limited $\pt/A$ acceptance. 
It is, therefore, important to investigate whether a non-monotonic extrapolation factor can result from trivial physics. 
To this end, we first examine the $\Ratio$ ratio as a function of $\pt/A$ in our toy-model simulation.
This is shown in Fig.~\ref{fig:acc} left panel for various $\theta$ values. The falling and rising characteristic of the shape is determined by the $\pt$ distribution of Eq.~(\ref{eq:pt}). 
The shapes are similar for the various fluctuation magnitudes and the overall ratio increases with $\theta$ as expected.
The right panel of Fig.~\ref{fig:acc} shows the extrapolation factor $f$ as a function of $\Delta n_\rn$. 
(Note that the total yields in simulation are known of course, not from extrapolation, but we keep use of the term ``extrapolation factor".) Two fiducial ranges are depicted, $\pt/A=0.1$--0.2 and 0.3--0.4~\gevc. 
The extrapolation factor $f$  varies with $\Delta n_\rn$; however, the variation is relatively small, about 1\% for the $\pt/A=0.3$--0.4~\gevc\ range for $\theta=2$, a fluctuation magnitude of a factor of 2 of that of Poisson. 
Note that we have examined only relatively narrow $\pt/A$ ranges at small $\pt/A$ values because of statistics considerations. The fiducial $\pt/A$ ranges in the STAR analysis are relatively large and wide, so no direct comparisons can be made between our study thus far and the STAR results. 

\section{More Realistic Simulations}

\begin{figure}[hbt]
\includegraphics[width=0.5\linewidth]{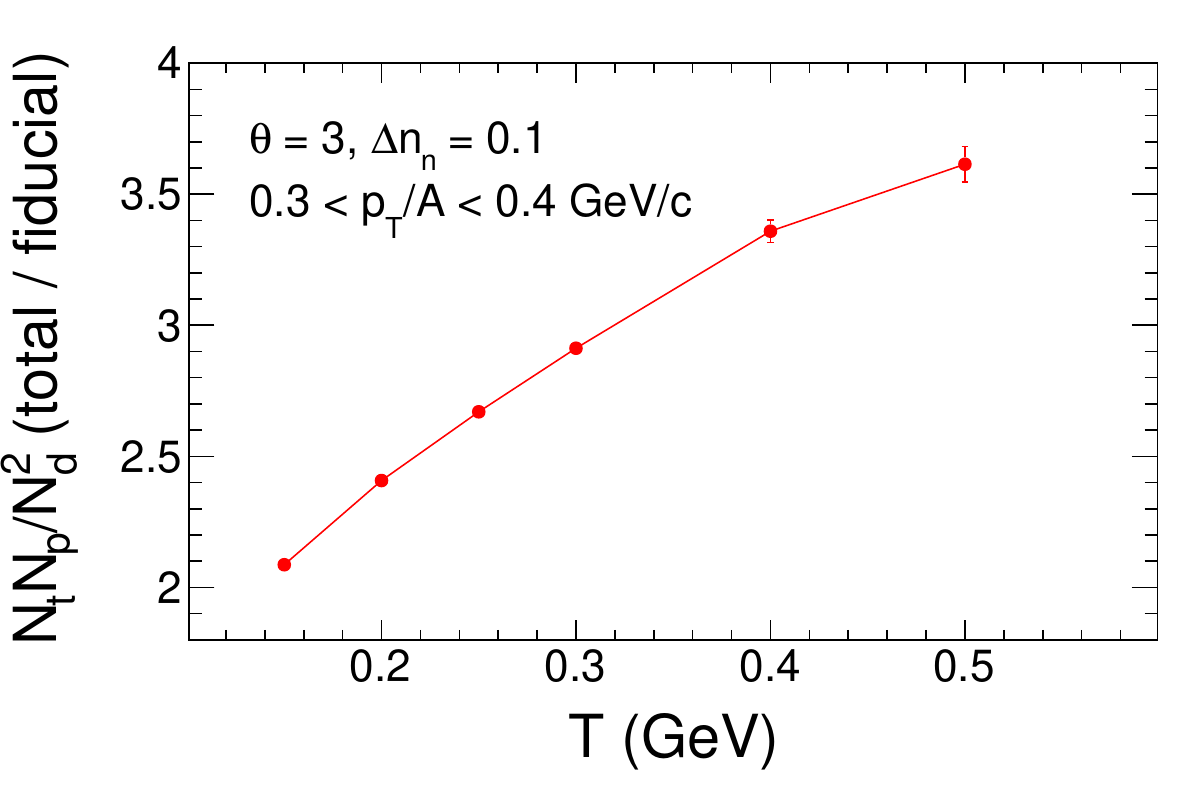}
\caption{The $\Ratio$ ``extrapolation factor" for the $\pt/A=0.3$--0.4~\gevc\ acceptance as a function of $T$, with $\theta=3$ or $\Delta n_\rn=0.1$. 
The light nuclei are formed by coalescence from a  system of average $\bar{N}_\rp=20$ protons and $\bar{N}_\rn=20$ neutrons randomly distributed within a cylinder of 10 fm radius and 10 fm length.} 
\label{fig:accT}
\end{figure}

In heavy ion collisions, the $\pt$ distributions can be significantly altered by collective radial flow. 
The shape of the $\pt$ distributions affects the $\Ratio$ ratio as functions of $\pt/A$. 
To investigate the effect of the nucleon $\pt$ spectral shape on the extrapolation factor for the $\Ratio$ ratio, we first simply vary the temperature parameter $T$ in Eq.~(\ref{eq:pt}) in our simulation. Figure~\ref{fig:accT} shows the $T$ dependence of the  extrapolation factor $f$ for $\pt/A=0.3$--0.4~\gevc, 
with a given neutron fluctuation value $\Delta n_\rn=0.1$ (i.e., $\theta=3$).  
The $f$ value varies with $T$ as expected, and the variation is monotonic. Thus, the $T$-dependent $\pt$ spectral shape would not cause an artificial non-monotonic $\Ratio$ enhancement from limited  to full acceptance in any given localized $T$ range.

The collective radial flow in heavy ion collisions creates a correlation between the momentum of a particle and its freeze-out radial position. 
This correlation is not present in the above simulations but can be important for coalescence.
In order to fully comprehend the implications of the STAR data~\cite{STAR:2022hbp}, in the following we perform simulations using more realistic kinematic distributions. We take the freeze-out information from measured  data~\cite{STAR:2017sal, STAR:2008med} as described below. The other simulation details are as same as those described in Sect.~\ref{sec}.


\begin{figure}[hbt]
    \includegraphics[width=0.48\linewidth]{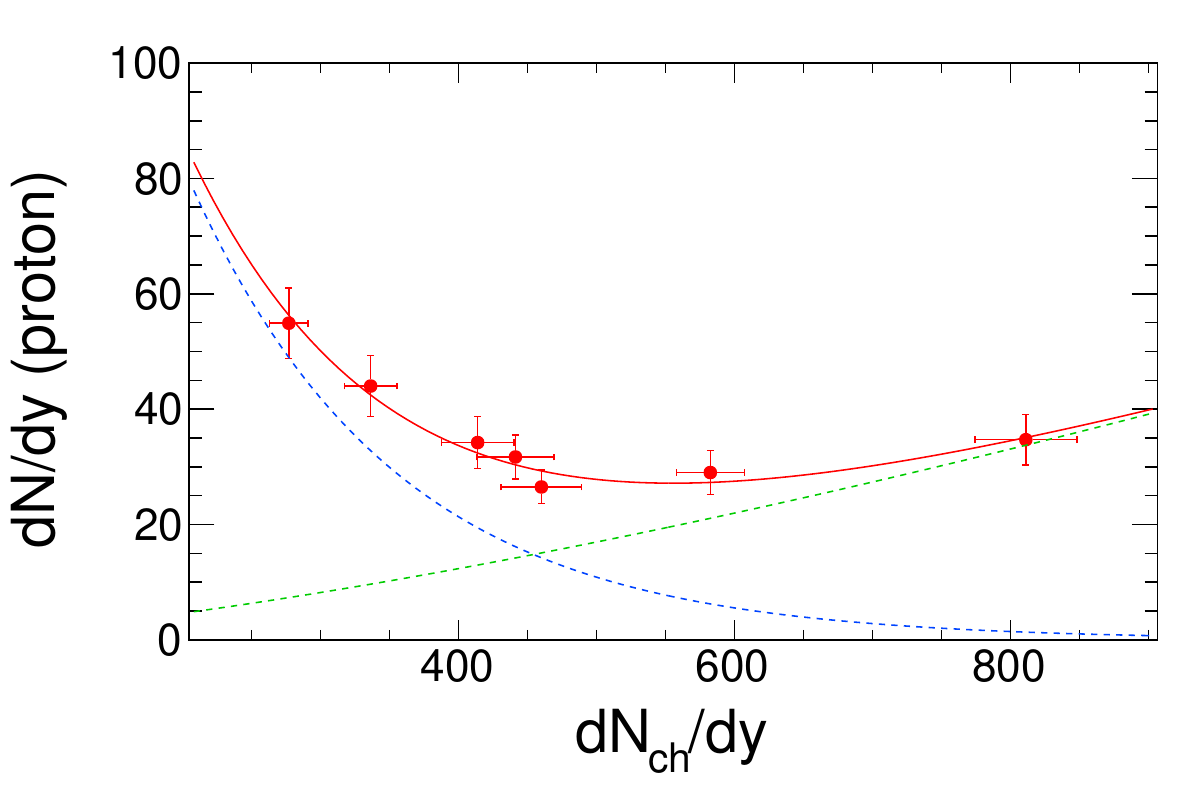}
    \includegraphics[width=0.48\linewidth]{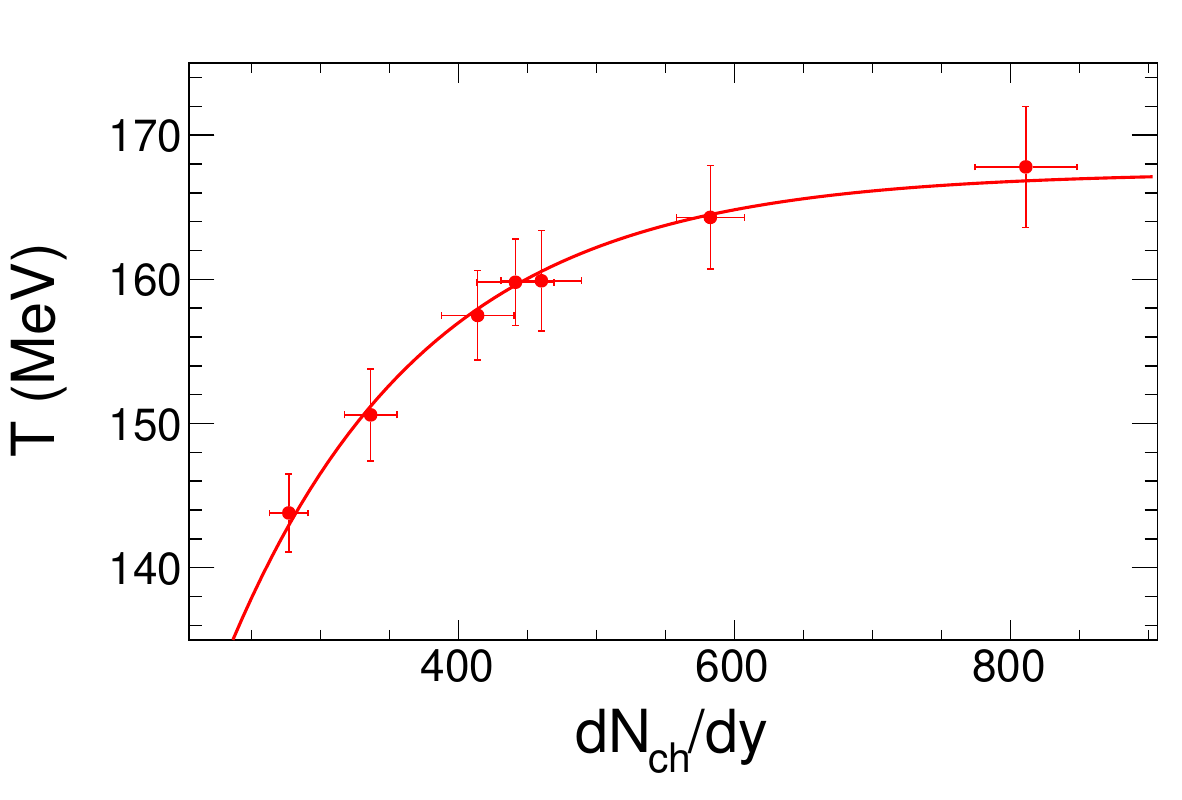}\hfill
    \includegraphics[width=0.48\linewidth]{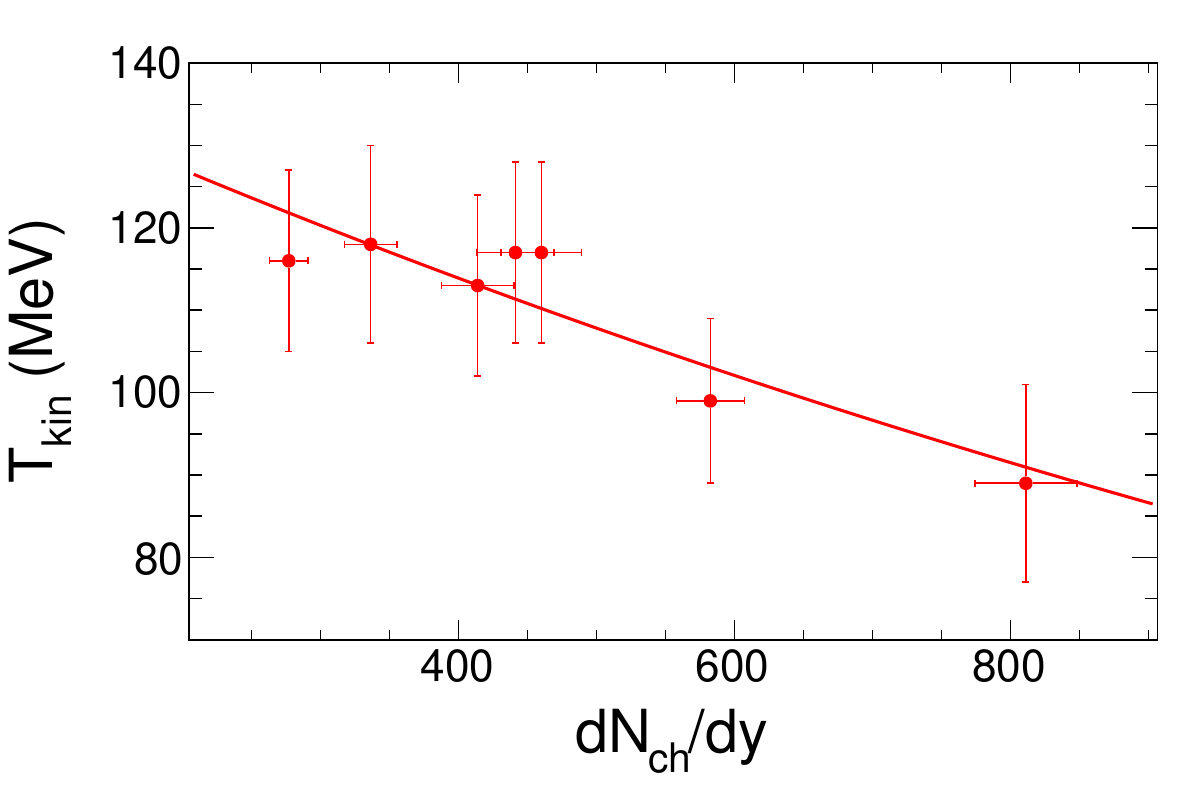}
    \includegraphics[width=0.48\linewidth]{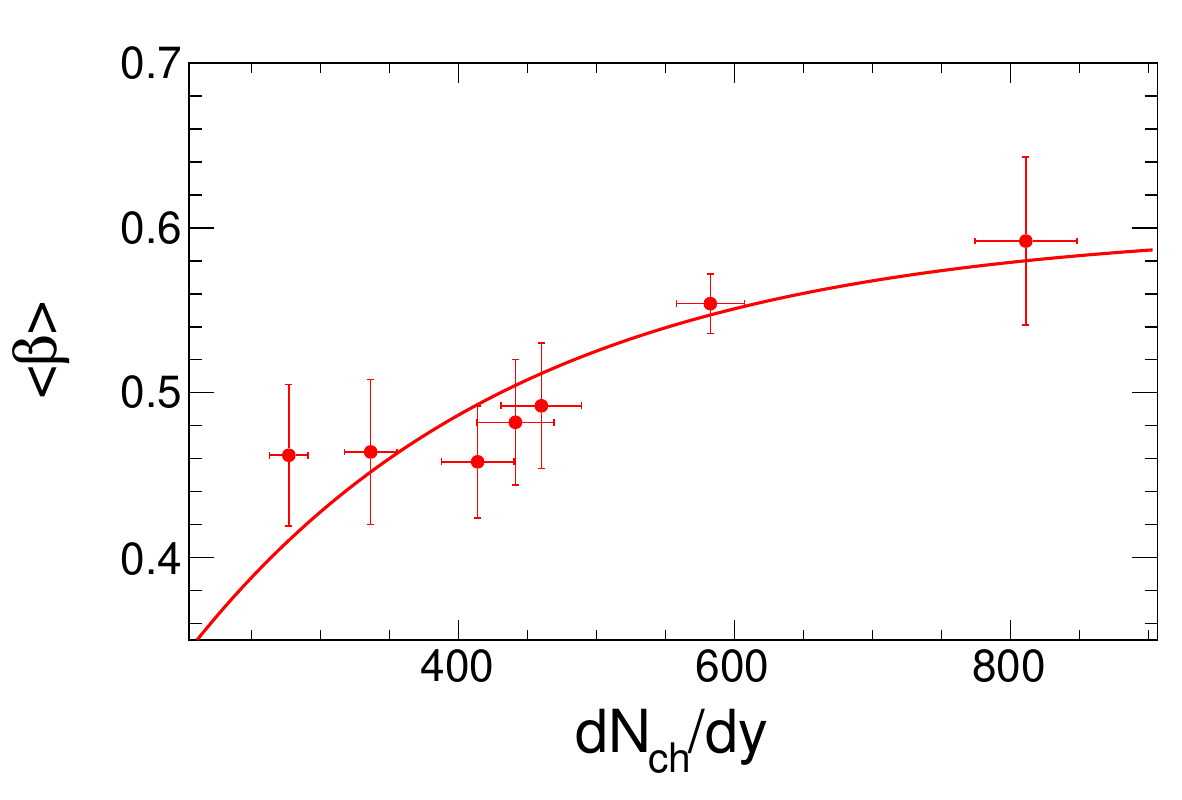}
    \caption{\label{fig:pars}
    The measured proton multiplicity density $dN_\rp/dy$ (upper left), the chemical freeze-out temperature $T_{\rm chem}$ (upper right), the kinetic freeze-out temperature $T_{\rm kin}$ (lower left), and the average radial velocity $\MEAN{\beta}$ (lower right) 
    as functions of the charged hadron multiplicity density $d\Nch/dy$~\cite{STAR:2017sal,STAR:2008med}. 
    The parameterizations to the data are superimposed and used as inputs to our simulations.
    } 
\end{figure}
The charged hadron mulitplicity $\Nch$, the proton multiplicity density $dN_\rp/dy$, the chemical freeze-out temperature $T_{\rm chem}$ and volume ($V_{\rm chem}\equiv\frac{4\pi}{3}R^3_{\rm sphere}$), the kinematic freeze-out temperature $T_{\rm kin}$ and the average collective radial flow velocity $\langle \beta \rangle$ in the Blast-Wave parameterization have been reported by the STAR experiment for various beam energies~\cite{STAR:2017sal, STAR:2008med}. 
We concentrate only on central 0-5\% collisions, and parameterize those freeze-out quantities as functions of $N_{\rm ch}$. The parameters have fit uncertainties, however, since we are only interested in input values into our simulation to study light nuclei coalescence, we take simply the parameterized values.
\begin{itemize}
\item The measured $dN_\rp/dy$ is shown in Fig.~\ref{fig:pars} (upper left) as a function of $N_{\rm ch}$. The measured protons come from two sources: transport protons whose abundance decreases with energy and produced protons whose abundance increases with energy. We thus parameterize $dN_\rp/dy$ by combination of two functions, one decreasing and the other increasing with $N_{\rm ch}$. The parameterization is given by $dN_\rp/dy=318e^{-\Nch/148} + 68\Nch^{1.4}$, as superimposed.
\item The chemical freeze-out temperature $T_{\rm chem}$ is shown in Fig.~\ref{fig:pars} (upper right) as a function of $\Nch$. The $T_{\rm chem}$ increases with $\Nch$ and then saturates, so we parameterize it by $T_{\rm chem} = 167(1-e^{-\Nch/144})$~MeV.
  %
  \item The effective sphere radius for the chemical freeze-out volume appears linear in $\Nch^{1/3}$, so we parameterize it as $R_{\rm sphere}=3.1+0.42\Nch^{1/3}$. 
  We allow the intercept to be a free parameter as otherwise the fit is not as good.
\item The kinetic freeze-out temperature $T_{\rm kin}$ is shown in Fig.~\ref{fig:pars} (lower left) as a function of $\Nch$. It is found to decrease with $\Nch$, and parameterized by $T_{\rm kin}=142e^{-\Nch/1827}$~MeV.
\item The average radial flow velocity is shown in Fig.~\ref{fig:pars} (lower right) as a function of $\Nch$. It increases with $\Nch$ and appears to saturate at large $\Nch$, so we parameterize it as $\MEAN{\beta}=0.6(1-e^{-\Nch/241})$.
\end{itemize}

The information of nucleons to be input into the coalescence model are those at the kinetic freeze-out. The kinetic freeze-out volume $V_{\rm kin}$ is obtained by assuming the system expands adiabatically, so $V_{\rm kin}/V_{\rm chem} = (T_{\rm chem}/T_{\rm kin})^{3/2}$. 
In our simulation, we assume the collision zone to be a cylinder with radius $R=7$~fm, and the length of the cylinder is determined by $V_{\rm kin}/(\pi R^2)$. 
Protons and neutrons are positioned uniformly and randomly inside the cylinder.
Both of the mean numbers of protons and neutrons are assumed to be $\bar{N_\rp}=\bar{N_\rn}=2\times dN_\rp/dy$, over two units of rapidity. 
The protons and neutrons are first generated according to the thermal distribution at $T_{\rm kin}$,
\begin{equation}
    dN/d\pt\propto \pt \mt e^{-\mt / T}\,.
\end{equation}
The generated protons and neutrons are boosted radially with a boost velocity dependent of the nucleon's radial position ($r$) within the cylinder,
\begin{equation}
    \beta = \beta_s (r/R)^n\,,
\end{equation}
where the surface velocity is given by $\beta_s=(1+n/2)\MEAN{\beta} $. In this study, the parameter $n$ for all collision energies is set to be $1$.

\begin{figure}[hbt]
    \includegraphics[width=0.48\linewidth]{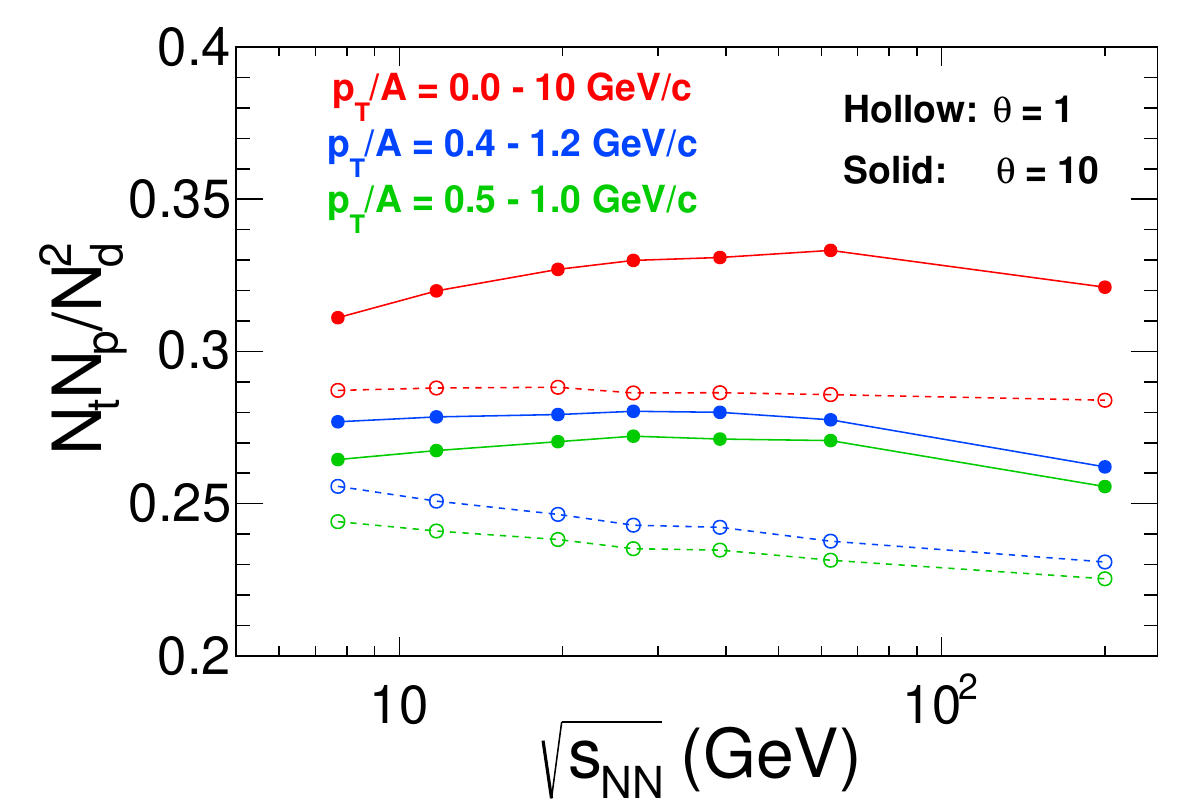}\hfill
    \includegraphics[width=0.48\linewidth]{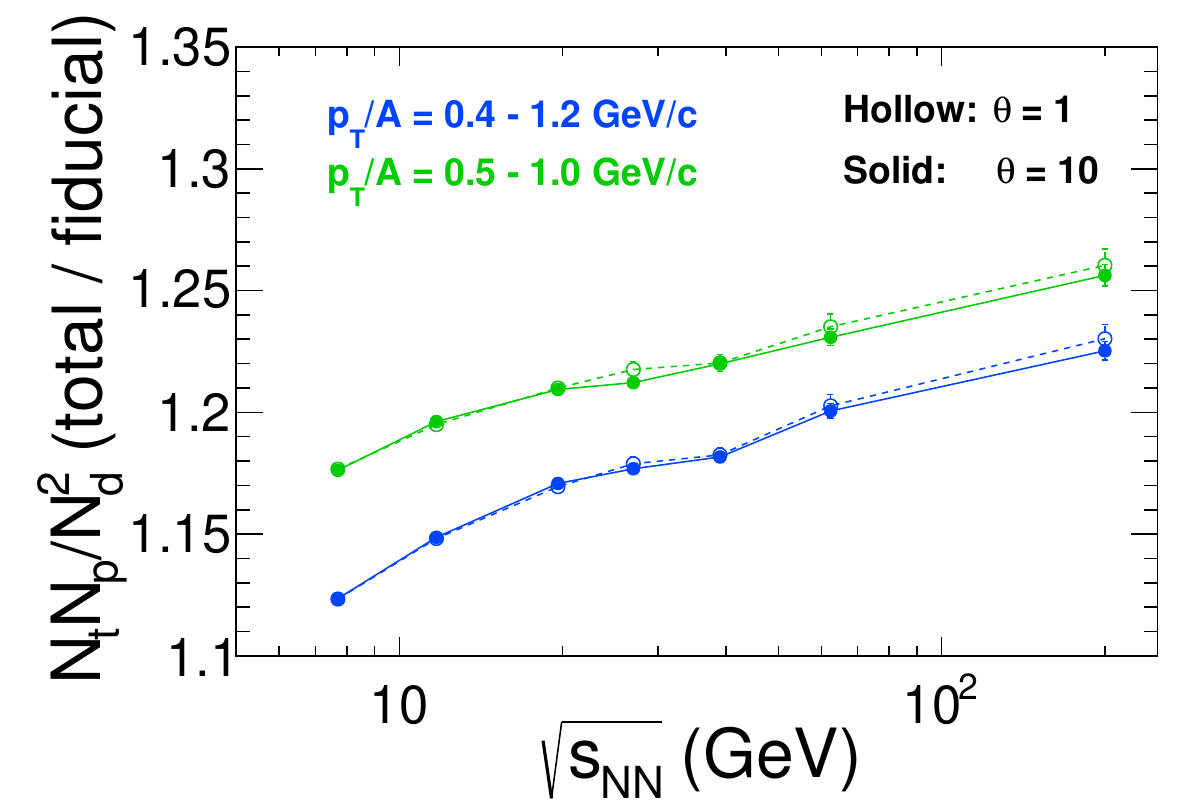}
    \caption{\label{fig:kin} (Left panel) The $\Ratio$ ratio as functions of $\snn$ for the entire $\pt$ range (red) and for two limited $\pt/A$ ranges (blue and green). Two values of $\theta$ are shown: $\theta=1$ for Poisson (hollow markers) and $\theta=10$ for enhanced neutron fluctuations (filled markers). (Right panel) The $\Ratio$ extrapolation factor for the $\pt/A=0.4$--1.2~\gevc\ (blue) and 0.5--1.0~\gevc\ (green) acceptance as functions of $\snn$, with $\theta=1$ (hollow markers) and $10$ (solid markers).}
\end{figure}

Figure~\ref{fig:kin} left panel shows the $\Ratio$ ratio calculated by our coalescence model as functions of $\snn$. The ratios are shown for the entire $\pt$ range as well as for two limited $\pt/A$ ranges measured by STAR, $0.4<\pt/A<1.2$~\gevc\ and $0.5<\pt/A<1.0$~\gevc. Results for two values of $\theta$ are depicted; $\theta=1$ corresponds to Poisson fluctuations in the neutron multiplicity, and  $\theta=10$ is for enhanced fluctuations with a magnitude corresponding to $\sim$10\% increase in the measured $\Ratio$ ratio by STAR (cf Eq.~(\ref{eq:dn}) where the typical $\bar{N}_\rn\sim60$--100 as shown in Fig.~\ref{fig:pars}). 
Similar to the STAR measurements~\cite{STAR:2022hbp}, the calculated $\Ratio$ ratios show weak energy dependence. The calculated $\Ratio$ values are somewhat lower than those measured.

The right panel of Fig.~\ref{fig:kin} shows the extrapolation factor $f$ of the $\Ratio$ ratio from $\pt/A=0.4$--1.2~\gevc\ and 0.5--1.0~\gevc, respectively. The extrapolation increases monotonically with increasing $\snn$ and is smooth. For a given $\pt/A$ range, the $f$ values are almost identical for the two $\theta$ values, and it is so for all $\theta$ values we have simulated. This suggests that $\pt$ extrapolation alone would not create, enhance, or reduce a local peak in the $\Ratio$ ratio as a function of $\snn$. In other words, the fluctuation increase in the $\snn\sim 20$--30~GeV region implied by the STAR extrapolated $\Ratio$ ratio~\cite{STAR:2022hbp} (such as a jump from the hollow red points to the solid red points at $\snn=20$--30~GeV in our simulation) should also be present in the ratios in the limited fiducial $\pt/A$ regions (a jump from the hollow blue/green points to the solid blue/green points). We therefore conclude that the local peak in the extrapolated $\Ratio$ ratio in the STAR measurements~\cite{STAR:2022hbp} is unlikely caused by physics.

\section{Summary}
We have simulated light nuclei production with the coalescence model from a system of nucleons with varying magnitude of neutron multiplicity fluctuations, $\Delta n_\rn$. It is found that the light nuclei yield ratio $\Ratio$ increases linearly with $\Delta n_\rn$, confirming the finding in Ref.~\cite{Sun:2017xrx,Sun:2018jhg}. We have further investigated the effect of finite acceptance by studying the ``extrapolation factor" $f$ for the $\Ratio$ ratio as a function of $\Delta n_\rn$ and the $\pt$ spectra parameter $T$. The $f$ value is found to be monotonic as functions of $T$ and $\Delta n_\rn$; the variations in the latter are relatively small. 

We have also carried out the coalescence model study with realistic kinematic distributions of nucleons using measured freeze-out parameters, the kinetic freeze-out temperature and collective radial flow velocity. The $\Ratio$ ratio is calculated by the coalescence model as functions of beam energy, and a weak beam energy dependence is found similar to experimental data. The extrapolation factor $f$ is found to be smooth and monotonic in beam energy and independent of the magnitude of neutron density fluctuations. We conclude that the extrapolation of the $\Ratio$ ratio in $\pt$ would not create, enhance, or reduce a local peak of the ratio in beam energy, and therefore the measured enhancement in the $\pt$-extrapolated $\Ratio$ ratio~\cite{STAR:2022hbp} is unlikely caused by physics.

Our study provides a necessary benchmark for light nuclei ratios as a probe of nucleon fluctuations, an important observable in the search for the critical point of nuclear matter. 
We have assumed in the present study that the fluctuations in the numbers of protons and neutrons are independent. One may implement varying degrees of correlations between those fluctuations and study the effect on the $\Ratio$ ratio. We leave such a study to a future work.


\section*{Acknowledgement}
We thank Dr.~Fuqiang Wang for suggesting the project and for many fruitful discussions.
This work was supported in part by the U.S.~Department of Energy (Grant No.~DE-SC0012910).

\bibliography{sample}
\end{document}